\documentclass[12pt]{article}

\usepackage{graphicx}
\usepackage{epsfig}
\usepackage{amssymb}

\normalbaselines

\def\lsim{\mathrel{\raise3pt\hbox to 8pt{\raise -6pt\hbox{$\sim$}\hss {$<$}}}} 
\newcommand{\be}{\begin{equation}}
\newcommand{\ee}{\end{equation}}
\newcommand{\bea}{\begin{eqnarray}}
\newcommand{\eea}{\end{eqnarray}}

\begin{document}


\noindent  

\begin{center}
{\Large \bf   Earth Flyby Anomalies    }\\[7mm]
{\bf Michael Martin Nieto\footnote{
Theoretical Division (MS-B285), Los Alamos National Laboratory, 
Los Alamos, New Mexico 87545 U.S.A.  {\it email:  mmn@lanl.gov}}
and 
John D. Anderson\footnote{
Jet Propulsion Laboratory, California Institute of  Technology,
Pasadena, CA 91109, U.S.A.  {\it email: jdandy@earthlink.net}  }
}\\
~\\
Michael Martin Nieto is a fellow at Los Alamos National Laboratory and \\
John D. Anderson is a senior research scientist emeritus at the Jet Propulsion 
Laboratory.

\end{center}

\vspace{2mm}

\begin{abstract}

In a reference frame fixed to the solar system's center of mass, a satellite's energy will change as it is deflected by a planet. But a number of satellites flying by Earth have also experienced energy changes in the Earth-centered frame -- and that's a mystery.\\

\end{abstract}



Planetary flybys began to receive much attention during the 1960s, when NASA's Jet Propulsion Laboratory (JPL) first started thinking about the would later become the Pioneer and Voyager missions -- the grand tours of the solar system conducted during the 1970s and the 1980s. The concept was to have the gravitational force of a planet give a “gravitational assist” to a satellite moving under the Sun's influence, so as to change the satellite’s direction and speed as measured in a barycentric reference frame tethered to the center of mass of the solar system. At the time, many found it surprising that a planet could transfer energy to a spacecraft, even though Joseph Lagrange, Carl Jacobi, and Felix Tisserand had long since established that the energies of comets could be affected by passing near Jupiter. After all, they might have reasoned, isn't the energy of a satellite flying through the solar system conserved?


\section{Planetary flybys}
\label{intro}

\begin{figure}[h!]
 \begin{center}
\noindent    
\includegraphics[width=3.4in]{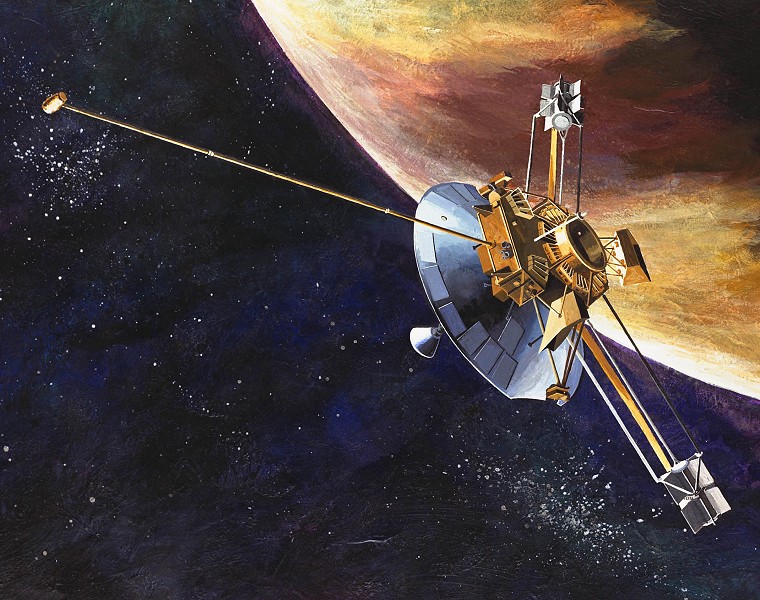} 
\end{center}
\caption{  Pioneer 10 and Jupiter, in an artist's rendition. In December 1973, Pioneer 10 was whipped around the giant planet and gained enough kinetic energy to escape the solar system. (Courtesy of NASA/JPL.)}
\label{pioatJ}
\end{figure}

When a third body such as a planet is considered, the energy of the satellite need not be conserved; only the total energy of the entire system must be constant. A flyby (see Figure \ref{pioatJ} ) can give kinetic energy to a spacecraft to boost its orbital velocity or even unbind it from the Sun's gravitational tug, or it can take away kinetic energy to slow the craft down for an inner-body encounter.


\begin{figure}[h!]
 \begin{center}
\noindent    
\includegraphics[width=3.4in]{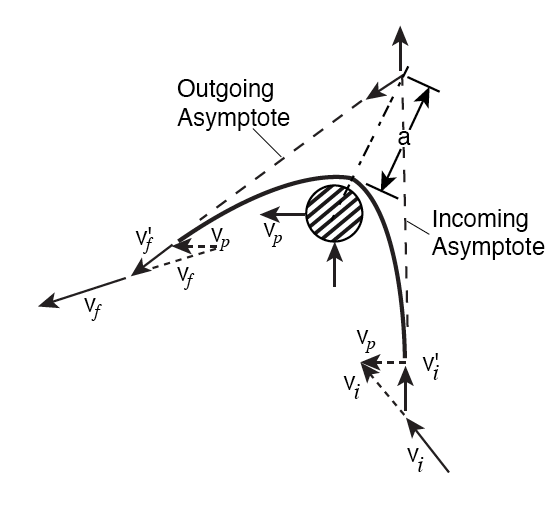}  
\end{center}
\caption{ Geometry and energetics of a flyby. In this sketch of a satellite's hyperbolic trajectory around a planet, the primed initial and final velocities are measured in the planet-centered frame, and unprimed velocities are measured in the barycentric frame. (Adapted from G. Flandro, Acta Astronaut. {\bf 12}, 329 (1966).)}   
\label{flandrofig}
\end{figure}

Figure \ref{flandrofig} describes a two-body system -- say, a planet and a satellite -- orbiting the Sun. 
The initial and final velocities 
in the heliocentric system are ${\bf v}_i$ and ${\bf v}_f$.    The initial and 
final velocities in the planetary system are ${\bf v}'_i$ and ${\bf v}'_f$.  The 
velocity of the planet in the solar system is ${\bf v}_p$.  The change in 
kinetic energy per unit mass is 
\begin{equation}
\Delta {\cal K} = ({\bf v}_f \cdot {\bf v}_f - {\bf v}_i\cdot {\bf v}_i)/2.  
\label{fly1}
\end{equation}
A little algebra gives one
\begin{equation}
\Delta {\cal K} = {\bf v}_p \cdot ({\bf v}_f' - {\bf v}_i').                 
\label{fly2}
\end{equation} 


Assume, for argument’s sake, that the planet is in a counterclockwise orbit around the Sun. Then, roughly speaking, if a satellite in the ecliptic (plane of the planet’s orbit) approaches from inside the orbit, travels behind the planet, and then goes around counterclockwise as illustrated in the figure, kinetic energy will be added to the spacecraft. Conversely, kinetic energy will be taken from the spacecraft that comes from inside the planetary orbit, travels in front of the planet, and goes around it clockwise.

Of course, the system does not violate the law of energy conservation. The energy change -- and, for that matter, angular-momentum change -- is absorbed by the planet. However, for that massive body, the relatively tiny change is not noticeable. In this Quick Study we’ll take a more detailed look at the Pioneer 10 flyby and then discuss a surprising result associated with Earth flybys: A number of craft have exhibited a small change in speed such that the outbound hyperbolic orbits in the Earth-centered system have a different energy than do the inbound orbits.


\section{Energy transfer in a Jupiter flyby}

The Pioneer spacecraft were the first to probe the major planets of the outer system and to go beyond those planets. They have now been overtaken by the faster Voyagers, which are approaching the edge of the solar system, where interstellar magnetic fields dominate. Pioneer 10, launched on 2 March 1972, was the first craft to reach a giant planet -- Jupiter, which it did on 4 December 1973. During its cruise from Earth to Jupiter, Pioneer 10 was bound to the solar system. With the Jupiter encounter, Pioneer 10 reached solar-system escape velocity.

\begin{figure}[h!]
\begin{center}
\noindent
\includegraphics[width=3.4in]{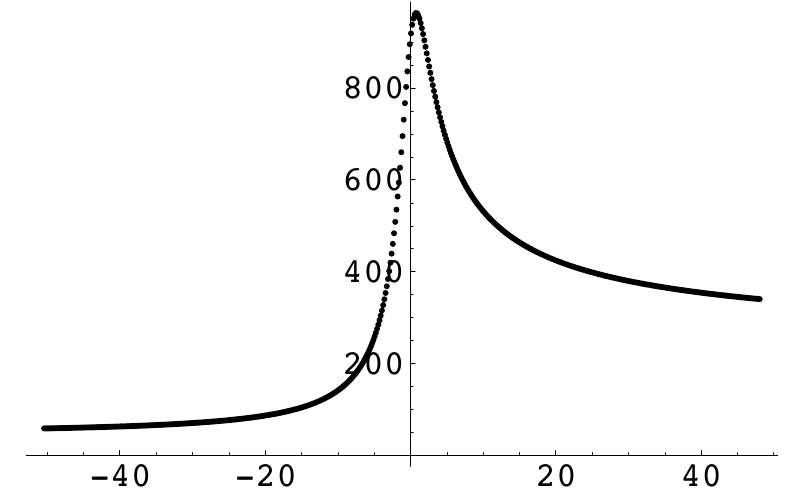}
\end{center}
\caption{  \small{This plot is of Pioneer 10's barycentric kinetic energy per unit mass 
$({\cal K})$
versus time. The zero of time represents the satellite’s closest approach to Jupiter.}
\label{10totE}} 
\end{figure}

Figure \ref{10totE}
illustrates the kinetic-energy transfer process during the Pioneer 10 flyby of Jupiter, as viewed in the barycentric frame. As Pioneer 10 approached Jupiter, it first gained a great deal of energy from the planet and then returned some back. Does that make sense?


Reconsider Figure \ref{flandrofig} and the dot-product expressions for $\Delta {\cal K}(t)$. Those equations are asymptotic formulae. But they can be recast as time-dependent equations if one substitutes an intermediate-time velocity ${\bf v}'(t)$ for the asymptotic velocity ${\bf v}'_f$. Then one has 
\begin{equation}
\Delta {\cal K}(t) = {\bf v}_p \cdot ({\bf v}'(t) - {\bf v}_i').
\end{equation}
The dot product with ${\bf v}_i'$ is a constant. The other dot product, as Figure \ref{10totE} shows, gives a spike just after Pioneer 10's closest approach to Jupiter. 
As the satellite goes around the back side of the planet, 
${\bf v}'(t)$ aligns with ${\bf v}_p$.  The maximum  of $\Delta {\cal 
K}$ is reached when ${\bf v}'(t)$ is parallel with ${\bf v}_p$.  This occurs 
just after perigee.  Then as the satellite swings around further, ${\bf v}'(t)$ 
goes out of alignment with ${\bf v}_p$ so $\Delta {\cal K}(t)$ decreases some.


\section{Anomalous Earrth flybys} 

The effect of an Earth flyby on a satellite depends on the bending of the spacecraft's geocentric trajectory. As discussed above, that change in direction can cause either a decrease or increase in the spacecraft’s energy as measured in the barycentric frame, depending on whether the spacecraft encounters Earth on the leading or trailing side of its orbital path. But, when measured from Earth, the spacecraft's energy should be the same before and after the flyby. The data indicate that such is not always the case.

Instead, for Earth flybys involving the Galileo, NEAR (Near Earth Asteroid Rendezvous), Cassini, Rosetta, and Messenger spacecraft, the final and initial energies were noticeably different. For example, Galileo 1's asymptotic final speed was $(4.13 \pm 0.03)$ mm/sec greater than it's asymptotic initial speed; other spacecraft had final speeds less than the initial speed. The changes, though typically only a millionth of the satellites' cruising speeds, were much too large to be explained in terms of any conventional time-explicit cause -- specifically, as arising from the nonsphericity of Earth or from the motions of the Moon and Sun. Morever, for the Rosetta flyby, JPL and European Space Agency programs independently determined the satellite's orbit. Thus the observed energy nonconservation is probably not due to code errors. So far, no one has identified a mechanism, either external or internal to the spacecraft, that can unambiguously produce the observed energy changes.

The mass distribution of Earth, and the dynamics of flybys are supposedly well known, yet the transfer of energy remains puzzling. Is there at least some phenomenological pattern to the anomalies? Indeed, a phenomenological formula exists that, perhaps fortuitously, fits the anomalies – at least for those spacecraft whose altitudes were below 2000 km:
\bea
\frac{\Delta v_\infty}{v_\infty} &=& K(\cos \delta_i - \cos \delta_f), 
     \label{formula} \\
K &=& \frac{2\omega_{\oplus} R_{\oplus}}{c} = 3.099 \times 10^{-6}.
\eea   
$\delta_i$ and $\delta_f$ are the initial (ingoing) and final (outgoing) 
declination angles, 
analogous to Earth's latitude but measured on the celestial sphere; $\omega_\oplus$ is Earth's angular velocity of rotation; $R_\oplus$ is its radius; and $c$ is the speed of light. With additional data, one might be able to generate a formula that also includes spacecraft achieving altitudes greater than 2000 km. But in any case, a phenomenological fit is not an explanation. What is the speed of light doing there? How does the declination produce a physical effect?

Given the anomalous energy transfer in Earth flybys, one might consider returning to the flybys of Jupiter and Saturn and looking for unexpected energy transfers there. Unfortunately, the uncertainties in the giant planets' gravitational fields make that program untenable. In addition, given the speed of a flyby trajectory, detection of the relatively small anomalous speed changes would require that the spacecraft be free of unknown, systematic accelerations. Such accelerations can be generated, for example, by onboard systems or by atmospheric drag. To be useful for analyzing flybys, the Doppler-shifted data sent by the satellite must be of high quality and referenced to ground-based atomic frequency standards. That requirement limits the number of candidates for further flyby tests, whether of Earth or other planets. However, given the importance of the gravity-assist technique as a means for conserving rocket fuel, flybys will occur in the future, and we feel it is worth the effort to overcome the difficulties just discussed.

Several physicists have proposed explanations of the Earth flyby anomalies. The least revolutionary invokes dark matter bound to Earth. Others include modifications of special relativity, of general relativity, or of the notion of inertia; a light speed anomaly; or an anisotropy in the gravitational field -- all of those, of course, deny concepts that have been well tested. And none of them have made comprehensive, precise predictions of Earth flyby effects. For now the anomalous energy changes observed in Earth flybys remain a puzzle. Are they the result of imperfect understandings of conventional physics and experimental systems, or are they the harbingers of exciting new physics? \\


{\bf ADDITIONAL RESOURCES} 

\begin{itemize}

\item J. A. Van Allen, 
``Gravitational Assist in Celestial Mechanics - a Tutorial," 
Am. J. Phys. {\bf 71}, 448-451 (2003). 

\item J. D. Anderson, J. K. Campbell, and M. M. Nieto,  
``The Energy Transfer Process in Planetary Flybys,"
New Astron. {\bf 12}, 383-397 (2007).
ArXiv: astro-ph/0608087.  

\item  J. D. Anderson, J. K. Campbell, J. F. Ekelund, J. Ellis, and J. E. 
Jordon, 
``Anomalous Orbital-Energy Changes Observed during Spacecraft Flybys of 
Earth,''
Phys. Rev. Lett. {\bf 100}, 091102/1-4 (2008).

\end{itemize}


\end{document}